# Hamiltonian thermodynamics

**Sergey A. Rashkovskiy**

*Ishlinsky Institute for Problems in Mechanics of the Russian Academy of Sciences, Vernadskogo Ave., 101/1, Moscow, 119526, Russia*

*Tomsk State University, 36 Lenina Avenue, Tomsk, 634050, Russia*

*E-mail: rash@ipmnet.ru, Tel. +7 906 0318854*

**Abstract.** It is believed that thermodynamic laws are associated with random processes occurring in the system and, therefore, deterministic mechanical systems cannot be described within the framework of the thermodynamic approach. In this paper, we show that thermodynamics (or, more precisely, a thermodynamically-like description) can be constructed even for deterministic Hamiltonian systems, for example, systems with only one degree of freedom. We show that for such systems it is possible to introduce analogs of thermal energy, temperature, entropy, Helmholtz free energy, etc., which are related to each other by the usual thermodynamic relations. For the considered Hamiltonian systems, the first and second laws of thermodynamics are rigorously derived, which have the same form as in ordinary (molecular) thermodynamics. It is shown that for Hamiltonian systems it is possible to introduce the concepts of a thermodynamic state, a thermodynamic process, and thermodynamic cycles, in particular, the Carnot cycle, which are described by the same relations as their usual thermodynamic analogs.

## 1 Introduction

Thermodynamics is undoubtedly one of the most successful physical theories. It arose and developed mainly as a phenomenological theory that generalizes experimentally observed facts and laws. It was within this approach that the basic (first, second and third) laws of thermodynamics were discovered, its basic concepts, such as energy, heat, temperature, entropy, etc., were defined. Thermodynamics has provided a unified approach to describing a wide class of processes associated with temperature changes in a variety of systems (atomic, molecular, plasma).

Only with the development of molecular-atomistic concepts and the emergence of statistical mechanics, thermodynamics did receive a physical (statistical) justification as a result of the movement and interaction of a large number of particles - atoms, molecules, ions.

At present, it is traditionally believed that the thermodynamic laws are based on statistical laws, that describe large ensembles of particles and, therefore, deterministic mechanical systems cannot be described within the framework of the thermodynamic approach.

In this paper, we show that thermodynamics (or, more precisely, a thermodynamically-like description) can be constructed even for deterministic Hamiltonian systems, for example, systems with only one degree of freedom.

In particular, we will show that for such systems, it is possible to introduce analogs of thermal energy, temperature, entropy, Helmholtz free energy, etc., which are related to each other by the usual thermodynamic relations. We will show that for near-Hamiltonian systems, the first and second laws of thermodynamics can be rigorously derived, which have the same form as in ordinary (molecular) thermodynamics. We will show that for Hamiltonian systems it is possible to introduce the concepts of thermodynamic state, thermodynamic process, and thermodynamic cycles, in particular, the Carnot cycle, which are described by the same relations as their usual thermodynamic analogs.

In order to distinguish the thermodynamics of Hamiltonian systems which is developed in this work from ordinary (traditional) thermodynamics, we will call it "Hamiltonian thermodynamics", while we will conventionally call ordinary thermodynamics "molecular thermodynamics".

## 2 First and second laws of thermodynamics for one-dimensional near-Hamiltonian systems

In this Section, we will follow the narrative [1].

Consider a one-dimensional finite motion of a Hamiltonian system that depends on some parameters $A = (a_1, \ldots, a_L)$, which we will call the external parameters of the system. The coordinate $q$, momentum $p$, energy $E$ of the system, etc. will be called the internal parameters of the system.

Then the Hamilton function of the system under consideration is $H = H(p, q, A)$.

We assume that with constant values of external parameters $A$ and in the absence of additional external influences, the system performs periodic motion with constant energy $E$ and a well-defined period $T(E, A)$.

Suppose that the parameters $A$ under the influence of some external reasons change slowly, so that

$$T \left|\frac{dA}{dt}\right| \ll |A| \qquad (1)$$

In classical mechanics, such a change in external parameters is called adiabatic [1,2], meaning by this only a slow change, and in no way connecting it with the concept of adiabaticity, which is introduced in thermodynamics.

With variable parameters $A$, the system is not closed and its energy is not conserved. However, due to condition (1), the rate $\dot{E}$ of the change in the energy of the system is also small. If we average this rate over the period $T$, and thereby smooth the "fast" fluctuations in its value, then the resulting value $\overline{\dot{E}} = \frac{d\overline{E}}{dt}$ will determine the rate of systematic slow change in the energy of the system. This rate is proportional to the rate $\dot{A}$ of the change in the parameters $A$.

Moreover, additional (including non-Hamiltonian) forces can act on the system under consideration, which also change the energy of the system.

In general, we can write

$$\frac{dE}{dt} = \frac{\partial H}{\partial t} + W \tag{2}$$

where $W$ is the power of additional (including non-Hamiltonian) forces acting on the system.

Taking into account that $\frac{\partial H}{\partial t} = \sum_{s=1}^{L} \frac{\partial H}{\partial a_s} \frac{da_s}{dt}$, one obtains

$$\frac{dE}{dt} = \sum_{s=1}^{L} \frac{\partial H}{\partial a_s} \frac{da_s}{dt} + W \tag{3}$$

The expression on the right-hand side of (3) depends not only on slowly varying external parameters $A$, but also on rapidly changing variables (internal parameters) $q$ and $p$.

To single out the systematic change in the energy of the system that interests us, it is necessary to average Eq. (3) over the period of fast motions. In this case, due to the slow change in the parameters $A$ (and, therefore, $\dot{A}$), we can move $\dot{A}$ outside the averaging sign.

As a result, we obtain

$$\frac{d\overline{E}}{dt} = \sum_{s=1}^{L} \frac{\partial \overline{H}}{\partial a_s} \frac{da_s}{dt} + \overline{W} \tag{4}$$

where $\overline{H} = \overline{E}$ means the averaging of the Hamilton function over the period of motion of the system at constant parameters $A$ and $W = 0$. The mean value $\overline{W}$ of the power of additional non-Hamiltonian forces acting on the system is also determined at constant values of the external parameters $A$.

We assume that $\overline{W}$ also satisfies the condition of a slow change in the energy of the system:

$$T|\overline{W}| \ll \overline{E} \tag{5}$$

In an explicit form, the averaging condition is as follows

$$\frac{\partial \overline{H}}{\partial a_s} = \frac{1}{T} \int_0^T \frac{\partial H}{\partial a_s} dt$$

Taking into account the Hamilton equation $\dot{q} = \frac{\partial H}{\partial p}$, we change the variables $dt = \frac{dq}{\partial H/\partial p}$.

As a result, one obtains

$$T = \int_0^T dt = \oint \frac{dq}{\partial H/\partial p} \tag{6}$$

and

$$\frac{\partial \overline{H}}{\partial a_s} = \frac{\oint \frac{\partial H/\partial a_s}{\partial H/\partial p} dq}{\oint \frac{dq}{\partial H/\partial p}} \qquad (7)$$

where the integration $\oint(...)dq$ is taken over one cycle of the oscillation in time at constant values of the parameters $A$ and $W = 0$ (i.e., in the absence of additional non-Hamiltonian influences on the system). Along such a path, the Hamilton function retains a constant value $E$, while the momentum $p$ is a definite function of the variable coordinate $q$ and constant independent parameters $E$ and $A$: $p = p(q, E, A)$. Then differentiating the equality $H(p, q, A) = E$ with respect to the parameters $a_s$, we obtain

$$\frac{\partial H}{\partial a_s} + \frac{\partial H}{\partial p}\frac{\partial p}{\partial a_s} = 0 \quad \text{or} \quad \frac{\partial H/\partial a_s}{\partial H/\partial p} = -\frac{\partial p}{\partial a_s}$$

Substituting this expression into the upper integral in (7) and writing the integrand in the lower integral in the form $\partial p/\partial E$, we rewrite Eq. (7) as

$$\frac{\partial \overline{H}}{\partial a_s} = -\frac{\oint \frac{\partial p}{\partial a_s} dq}{\oint \frac{\partial p}{\partial E} dq} \qquad (8)$$

We introduce the action integral

$$J = \frac{1}{2\pi} \oint p\, dq \qquad (9)$$

where the integral is also taken over one cycle of the oscillation in time at constant values of $E$ and $A$, as well as at $W = 0$. This means that the action integral $J$ is a function of the parameters $E$ and $A$:

$$J = J(E, A) \qquad (10)$$

Then relation (8) takes the form

$$\frac{\partial \overline{H}}{\partial a_s} = -\frac{\left(\frac{\partial J}{\partial a_s}\right)_E}{\left(\frac{\partial J}{\partial E}\right)_A} \qquad (11)$$

Here, the notation $\left(\frac{\partial F}{\partial x}\right)_y$, accepted in thermodynamics [3], is used, which means that the derivative of the function $F$ is taken with respect to the parameter $x$ at a constant value of the parameter $y$.

Taking into account (11), we rewrite equation (4) in the form

$$\delta Q = dE + \sum_{s=1}^{L} B_s\, da_s \qquad (12)$$

where we introduced the notations

$$B_s = \tau \left(\frac{\partial J}{\partial a_s}\right)_E \qquad (13)$$

$$\frac{1}{\tau} = \left(\frac{\partial J}{\partial E}\right)_A \qquad (14)$$

$$\delta Q = \overline{W} dt \qquad (15)$$

Hereinafter, the sign of averaging over the energy is omitted, and $E$ means the energy of the system averaged over the period $T$.

Equation (12) describes an elementary process that occurs in a short time $dt$. It should be borne in mind here that the motion of a near-Hamiltonian system has two characteristic time scales: the oscillation period of the Hamiltonian system $T$ and the characteristic time $T_A$ of the change in external parameters $A$ (we assume that the characteristic time of changes in the system associated with the action of non-Hamiltonian forces, has the same order as $T_A$). Due to conditions (1) and (5), $T_A \gg T$. After averaging the parameters of the system (in particular, energy) over the oscillation period $T$, the time has a characteristic scale only $T_A$. In particular, the derivative $\frac{d\overline{E}}{dt}$ depends on time with the scale $T_A$.

Then, when considering the processes associated with a change in external parameters or with a non-Hamiltonian impact on the system, the time intervals $dt$ satisfying the condition $T \ll dt \ll T_A$ are considered as elementary time intervals.

Note that since the power $\overline{W}$ is non-Hamiltonian, the quantity $\delta Q$ in the general case cannot be reduced to the differential of any function; therefore, instead of the sign of the differential "$d$", we use here the variation sign "$\delta$".

Relation (12) up to notations corresponds to the first law of thermodynamics and has the same physical meaning - the law of conservation of energy. In this case, the components $B_s da_s$ describe the mechanical work that the system performs when the external parameters $A$ change, and the quantity $\delta Q$ describes the energy that the system receives from external non-Hamiltonian sources at constant values of the external parameters $A$. In thermodynamics, the quantity $\delta Q$ is called the amount of heat or thermal energy (or just heat). We will keep this term for Hamiltonian systems as applied to the quantity $\delta Q$ (15), sometimes adding the adjective "Hamiltonian": Hamiltonian heat. Further we will see that this does not contradict the thermodynamic definition of heat [3].

Taking into account (10), we can write
$$dJ = \left(\frac{\partial J}{\partial E}\right)_A dE + \sum_{s=1}^{L} \left(\frac{\partial J}{\partial a_s}\right)_E da_s$$
or, using the notation (13) and (14), one obtains
$$\tau dJ = dE + \sum_{s=1}^{L} B_s da_s \tag{16}$$
Comparing (12) and (16), we obtain
$$\delta Q = \tau dJ \tag{17}$$
Taking into account the meaning of $\delta Q$ and relation (14), we come to the conclusion that the action integral $J$ for the systems under consideration plays the same role as entropy in ordinary thermodynamic systems, while the parameter $\tau$ plays the role of temperature. From this point of view, relation (14) is the usual thermodynamic definition of temperature [3]. For this reason, the parameters $J$ and $\tau$ will also be called the entropy and temperature of the Hamiltonian system (Hamiltonian entropy and Hamiltonian temperature). In order not to confuse the Hamiltonian temperature with the period $T$, and the Hamiltonian entropy with the action $S$, we will further use the introduced notations $J$ and $\tau$.

Thus, equations (16) and (17) represent the second law of thermodynamics for near-Hamiltonian systems.

We see that the introduced concepts of temperature, entropy, and heat, as well as the second law of thermodynamics (17) for near-Hamiltonian systems are in no way connected with random influences or with internal random processes in a Hamiltonian system.

It is easy to establish the physical meaning of the Hamiltonian temperature $\tau$. Indeed, differentiating (9) with respect to $E$ at constant external parameters $A$ and taking into account (6), we obtain

$$\left(\frac{\partial J}{\partial E}\right)_A = \frac{1}{2\pi}\oint \left(\frac{\partial p}{\partial E}\right)_A dq = \frac{T}{2\pi} \tag{18}$$

Taking into account (14), one obtains

$$\tau = \omega \tag{19}$$

where $\omega = 2\pi/T$ is the radian frequency of system oscillation.

Thus, the temperature of a near-Hamiltonian system is equal to the radian frequency of its oscillation. Hence it follows that the higher the temperature (oscillation frequency) of the Hamiltonian system, the faster the internal motion in the system occurs. This is consistent with the physical meaning of temperature in ordinary thermostatistics: the higher the temperature, the faster the internal motions (of atoms and molecules) in a thermodynamic system occur. It is interesting to note that in ordinary thermodynamic systems, when the temperature tends to zero, any thermal motion stops. Taking into account (19), we see that for the considered Hamiltonian systems the limit with zero Hamiltonian temperature (i.e., with zero frequency and infinite period) corresponds, in fact, to the cessation of the internal motion of the system.

Note that, in the general case, any function $J' = \sigma(J)$ can be taken as the entropy of the Hamiltonian system; in this case, according to (14), the definition of temperature changes: $\frac{1}{\tau'} = \frac{1}{\tau}\frac{d\sigma}{dJ}$ and temperature $\tau'$ does not coincide with the radian frequency (19), while the generalized forces (13) do not change: $B'_s = B_s$. With a new choice of entropy, $\tau' dJ' = \tau dJ$, which means that the first and second laws of thermodynamics (12) and (17) do not change. In a particular case, the entropy and temperature of the Hamiltonian system can be redefined using the constant factors $k_J$ and $k_\tau$, which satisfy the condition $k_J k_\tau = 1$. This will only result in changing in the scales and units of entropy and temperature.

According to (17), the "thermal" impact on the Hamiltonian system is directly related to the change in the action integral $J$. Therefore, according to the thermodynamic definition, the adiabatic impact on the Hamiltonian system ($\delta Q = 0$) is an impact that does not change the action integral, even if the external parameters A change.

As is known, in mechanics [1, 4], the quantity $J$ is an "adiabatic invariant" in the sense that it does not change with a slow change in the external parameters of the system. The concept of "adiabatic impact" [1, 4] which is traditionally used in mechanics, means only a slow change in the external parameters of the system and is in no way connected with the thermodynamic definition of adiabaticity. The thermodynamics of near-Hamiltonian systems developed in this work introduces a strict definition of the concept of adiabaticity, which for ordinary thermodynamic systems turns into the usual thermodynamic definition. So we arrive at an amazing result: the concept of adiabatic invariance, introduced intuitively in mechanics, does indeed correspond to the strict thermodynamic definition of adiabaticity.

## 3 Equations of state and Helmholtz free energy of a Hamiltonian system

According to definitions (13) and (14) and, taking into account (10), we can write
$$B_s = B_s(\tau, E, A); \quad \tau = \tau(E, A)$$
From this we can obtain the relations
$$E = E(\tau, A); \quad B_s = B_s(\tau, A) \tag{20}$$
which can be called the equations of state of the Hamiltonian system. The first of relations (20) can be called the energy equation of state, while the second - according to the thermodynamic tradition - the thermal equation of state.

Using the second law of thermodynamics for near-Hamiltonian systems in the form (16), it is easy to establish a connection between the thermal and energy equations of state [3].

Indeed, taking into account (20), we write (16) in the form
$$dJ = \frac{1}{\tau}\left(\frac{\partial E}{\partial \tau}\right)_A d\tau + \frac{1}{\tau}\sum_{s=1}^{L}\left(\left(\frac{\partial E}{\partial a_s}\right)_\tau + B_s\right) da_s \tag{21}$$

Relation (10) taking into account (20) can be written in the form $J = J(\tau, A)$, where the parameters $\tau$ and $A$ are considered independent. Then from equation (21) we obtain
$$\left(\frac{\partial J}{\partial \tau}\right)_A = \frac{1}{\tau}\left(\frac{\partial E}{\partial \tau}\right)_A; \quad \left(\frac{\partial J}{\partial a_s}\right)_\tau = \frac{1}{\tau}\left(\left(\frac{\partial E}{\partial a_s}\right)_\tau + B_s\right) \tag{22}$$

Equating the mixed derivatives $\frac{\partial^2 J}{\partial \tau \partial a_s}$ and $\frac{\partial^2 J}{\partial a_s \partial \tau}$, as well as $\frac{\partial^2 J}{\partial a_r \partial a_s}$ and $\frac{\partial^2 J}{\partial a_s \partial a_r}$, where $s \neq r$, we obtain

$$\tau \left(\frac{\partial B_s}{\partial \tau}\right)_A = \left(\frac{\partial E}{\partial a_s}\right)_\tau + B_s \qquad (23)$$

and

$$\frac{\partial B_s}{\partial a_r} = \frac{\partial B_r}{\partial a_s} \qquad (24)$$

The second law of thermodynamics in the form (16) can be rewritten as

$$dF = -J d\tau - \sum_{s=1}^{L} B_s da_s \qquad (25)$$

where we introduced the notation

$$F(\tau, A) = E - \tau J \qquad (26)$$

It follows from (25) that

$$J = -\left(\frac{\partial F}{\partial \tau}\right)_A, B_s = -\left(\frac{\partial F}{\partial a_s}\right)_\tau$$

To within the notations, function (26) coincides with the Helmholtz free energy in thermodynamics [3].

As in ordinary thermodynamics, other thermodynamic potentials can be introduced for the Hamiltonian system.

Consider the action for a Hamiltonian system, which is determined by the relation

$$S = \int_0^t L(q, \dot{q}, t) dt \qquad (27)$$

where

$$L(q, \dot{q}, t) = p\dot{q} - H(p, q, t) \qquad (28)$$

is the Lagrange function of the Hamiltonian system.

For a periodic system, action (27), (28) over one cycle of the oscillation in time is

$$S_T = \int_0^T L(q, \dot{q}) dt$$

or taking into account (28), one obtains

$$S_T = \oint p dq - ET \qquad (29)$$

Taking into account (9) and (19), we write relation (29) in the form

$$\frac{\tau}{2\pi} S_T = \tau J - E \qquad (30)$$

Comparing (30) with (26), one obtains

$$\frac{\tau}{2\pi} S_T = -F \qquad (31)$$

So the action of the Hamiltonian system over one cycle of the oscillation in time is expressed in terms of the Hamiltonian temperature and the Helmholtz free energy.

# 4 Non-quasi-static thermodynamic processes for Hamiltonian systems

The state of the Hamiltonian system with given external parameters and energy will be called the thermodynamic state, in contrast to the internal state of the Hamiltonian system, which is characterized by instantaneous values of momenta and coordinates $p, q$. Thus, the thermodynamic state of a Hamiltonian system is its periodic motion with constant external parameters and energy.

By a thermodynamic process in a Hamiltonian system, we will, as usual, mean a process in which the external parameters $A$ of the system change, and an additional (thermal) exchange of energy occurs between the Hamiltonian system and external systems. In this case, obviously, internal processes associated with periodic motion also occur in the Hamiltonian system, but these processes are not considered thermodynamic.

So far, we have considered quasi-static thermodynamic processes in a Hamiltonian system that satisfy conditions (1) and (5). Such processes are equilibrium and reversible. Indeed, under conditions (1) and (5), the periodic Hamiltonian system at each period behaves as if it had constant external parameters and energy, but from period to period the external parameters and energy can slowly change. Such a process can be carried out in the opposite direction, and in this case the system will go through the same thermodynamic states (i.e. the same values of external parameters, energy, temperature, etc.), but in the reverse sequence.

Let us now consider a non-quasi-static process in a Hamiltonian system when changes in external parameters and the supply of Hamiltonian heat do not satisfy conditions (1) and (5). Obviously, such a process will no longer be equilibrium and reversible.

Consider a Hamiltonian system, whose Hamiltonian function depends explicitly on external parameters $A$, which are functions of time:

$$H(p, q, t) = H(p, q, A(t)) \qquad (32)$$

Moreover, we will assume that additional impact are exerted on the system, as a result of which the system becomes non-Hamiltonian, although near-Hamiltonian.

In the general case, such a system is described by the equations

$$\dot{q} = \frac{\partial H}{\partial p} + V; \dot{p} = -\frac{\partial H}{\partial q} + F \qquad (33)$$

where $V$ and $F$ are given functions of parameters $p, q$ and $t$. The function $F$ is an additional (non-Hamiltonian) force, while the function $V$ describes the deviation from the Hamiltonian relation between momentum and velocity.

Taking into account (32) and (33), for the rate of change of the Hamilton function $\frac{dH}{dt} = \frac{\partial H}{\partial p}\dot{p} + \frac{\partial H}{\partial q}\dot{q} + \frac{\partial H}{\partial t}$ we obtain

$$\frac{dH}{dt} = \frac{\partial H}{\partial p}F + \frac{\partial H}{\partial q}V + \sum_{s=1}^{L}\frac{\partial H}{\partial a_s}\frac{da_s}{dt} \tag{34}$$

Comparing (34) with (3), we write explicitly the power of non-Hamiltonian forces:

$$W = \frac{\partial H}{\partial p}F + \frac{\partial H}{\partial q}V \tag{35}$$

The power of non-Hamiltonian forces averaged over the oscillation period of the Hamiltonian system is determined by the relation

$$\overline{W} = \frac{1}{T}\int_0^T \left(\frac{\partial H}{\partial p}F + \frac{\partial H}{\partial q}V\right)dt \tag{36}$$

or taking into account (33), one obtains

$$\overline{W} = \frac{1}{T}\oint(Fdq - Vdp) \tag{37}$$

where the integral is taken over one oscillation period of the Hamiltonian system.

Using (15), we write in an explicit form the expression for the amount of Hamiltonian heat received by the system in the time $dt \gg T$:

$$\delta Q = \frac{\tau dt}{2\pi}\oint(Fdq - Vdp) \tag{38}$$

where $\tau$ is the Hamiltonian temperature of system (19); the elementary time interval $dt$ is associated with slow changes in the system due to non-Hamiltonian impacts.

When deriving relation (38), we did not make any assumptions about the nature of the thermodynamic process; in particular, we did not assume that the process is quasi-static. Thus, relation (38) is valid for any thermodynamic processes: both quasi-static and non-quasi-static.

Let us now consider a non-quasi-static thermodynamic process involving a Hamiltonian system. The change in the energy of the Hamiltonian system due to the change in external parameters and non-Hamiltonian external impact is described by equation (3). Averaging Eq. (3) over the oscillation period of the Hamiltonian system, we obtain

$$\frac{dE}{dt} = \sum_{s=1}^{L}\overline{\frac{\partial H}{\partial a_s}\frac{da_s}{dt}} + \overline{W} \tag{39}$$

We take into account here that the thermodynamic process is not quasi-static, i.e. condition (1) is not satisfied. In this case

$$\overline{\frac{\partial H}{\partial a_s}\frac{da_s}{dt}} \neq \frac{\partial \overline{H}}{\partial a_s}\frac{d\overline{a_s}}{dt} \tag{40}$$

Let us consider an elementary non-quasi-static process in which the system passes from the thermodynamic state $(E, A)$ to the thermodynamic state $(E + dE, A + dA)$. In this case, Eq. (39) can be written as

$$dE = dt \sum_{s=1}^{L} \overline{\frac{\partial H}{\partial a_s} \frac{da_s}{dt}} + \delta Q \qquad (41)$$

where the Hamiltonian heat $\delta Q$ obtained by the system in the course of an elementary non-quasi-static thermodynamic process is determined by relation (15), (38), and the "slow" time $dt$ is still much longer than the oscillation period of the Hamiltonian system and much longer than the characteristic time of rapid changes in external parameters $A$.

Let us now consider a quasi-static process that occurs between the same thermodynamic states of the Hamiltonian system.

The second law of thermodynamics for a quasi-static process can be written in the form

$$\tau dJ = dE - \sum_{s=1}^{L} \frac{\partial \overline{H}}{\partial a_s} \frac{d\overline{a_s}}{dt} da_s \qquad (42)$$

where $dJ$ is the change in the Hamiltonian entropy of the system during its transition from the thermodynamic state $(E, A)$ to the thermodynamic state $(E + dE, A + dA)$ in the course of a quasi-static process; $\tau$ is the Hamiltonian temperature of the system in the thermodynamic state $(E, A)$.

Because both quasi-static and non-quasi-static processes occur between the same thermodynamic states of the Hamiltonian system, the change in energy $dE$ in equations (41) and (42) is the same. Then, substituting (41) into (42), we obtain

$$\tau dJ - \delta Q = dt \sum_{s=1}^{L} \left( \overline{\frac{\partial H}{\partial a_s} \frac{da_s}{dt}} - \frac{\partial \overline{H}}{\partial a_s} \frac{d\overline{a_s}}{dt} \right) \qquad (43)$$

where $\frac{d\overline{a_s}}{dt} dt = da_s$ is the change in the external parameters of the system in the course of an elementary quasi-static thermodynamic process.

Equation (43) is the second law of thermodynamics for quasi-static and non-quasi-static processes. In particular, for quasistatic processes, we have the condition

$$\overline{\frac{\partial H}{\partial a_s} \frac{da_s}{dt}} = \frac{\partial \overline{H}}{\partial a_s} \frac{d\overline{a_s}}{dt} \qquad (44)$$

and the right-hand side of equation (43) vanishes. In this case, the second law of thermodynamics takes the form (17). For non-quasi-static processes, condition (40) holds, and the right-hand side of equation (43) is not zero.

Obviously

$$\overline{\frac{\partial H}{\partial a_s} \frac{da_s}{dt}} - \frac{\partial \overline{H}}{\partial a_s} \frac{d\overline{a_s}}{dt} = \overline{\left( \frac{\partial H}{\partial a_s} - \frac{\partial \overline{H}}{\partial a_s} \right) \left( \frac{da_s}{dt} - \frac{d\overline{a_s}}{dt} \right)} \qquad (45)$$

Taking into account (45), we write equation (43) in the form

$$\tau dJ - \delta Q = dt \sum_{s=1}^{L} \overline{\left( \frac{\partial H}{\partial a_s} - \frac{\partial \overline{H}}{\partial a_s} \right) \left( \frac{da_s}{dt} - \frac{d\overline{a_s}}{dt} \right)} \qquad (46)$$

In contrast to ordinary thermodynamic processes (in molecular systems), for the considered Hamiltonian system, in the general case, we cannot say anything about the sign of the right-hand side of Eq. (46); therefore, we can only assert that for non-quasi-static processes in the Hamiltonian system

$$\tau dJ \neq \delta Q \tag{47}$$

Condition (47) can be considered as the second law of thermodynamics for non-quasi-static processes in a Hamiltonian system, which transforms into equation (17) in the case of quasi-static processes.

Thus, we see that, in the general case, for Hamiltonian systems, in contrast to ordinary molecular systems, the second law of thermodynamics does not indicate the direction of the nonequilibrium processes.

The analysis performed allows one to give a more rigorous definition of a quasi-static thermodynamic process than (1): a thermodynamic process can be called a quasi-static process if the condition (44) is satisfied.

## 5 Thermodynamic cycles for Hamiltonian systems

Of particular interest are thermodynamic cycles, i.e. thermodynamic processes in which the initial and final thermodynamic states of the Hamiltonian system are the same.

Using the first law of thermodynamics (12), for the thermodynamic cycle we obtain

$$\Delta Q = \sum_{s=1}^{L} \oint B_s da_s \tag{48}$$

where $\Delta Q = \oint \delta Q$ - the amount of heat received by the system for the entire thermodynamic cycle. Here, in contrast to the previous sections, the notation $\oint ...$ means the integral over the thermodynamic cycle, and it is taken into account that for a thermodynamic cycle, by definition, $\oint dE = 0$. Equation (48) shows that the work performed by the system over a thermodynamic cycle due to changes in external parameters is equal to the amount of Hamiltonian heat received by the system during the same cycle.

Using (17), for the thermodynamic cycle of the Hamiltonian system, we obtain the Clausius equality

$$\oint \frac{\delta Q}{\tau} = 0 \tag{49}$$

where we have taken into account that $\oint dJ = 0$.

For a Hamiltonian system, as for a conventional thermodynamic system, various thermodynamic cycles can be considered. The analysis of these cycles is not difficult, and we will not dwell on this. Consider just the most important thermodynamic cycle: the Carnot cycle. Similar to conventional thermodynamic system, the Carnot cycle for a Hamiltonian system consists of two isothermal ($\tau = \tau_1 = const$ and $\tau = \tau_2 = const, \tau_1 > \tau_2$) processes and two adiabatic (isentropic: $J = J_1 = const$ and $J = J_2 = const, J_1 > J_2$) processes (Fig. 1).

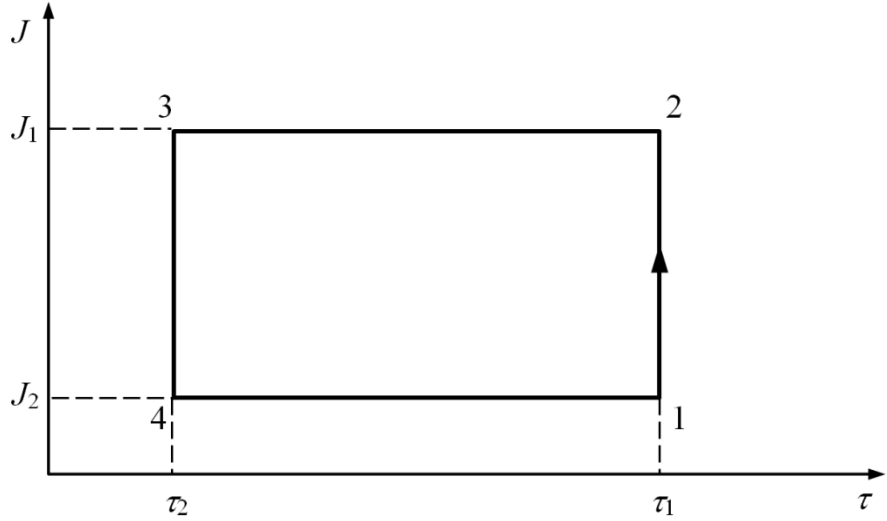

Fig. 1. The Carnot cycle for a Hamiltonian system. The arrow indicates the direction of the cycle.

According to (17), the system receives heat $\Delta Q_{12} = \tau_1(J_1 - J_2)$ in the process 1-2 and gives away (loses) heat $\Delta Q_{34} = \tau_2(J_1 - J_2)$ in the process 3-4. According to (48), the Hamiltonian system performs the work during the entire cycle

$$w = \sum_{s=1}^{L} \oint B_s da_s = \Delta Q \equiv \Delta Q_{12} - \Delta Q_{34} \tag{50}$$

or according to (17)

$$w = (\tau_1 - \tau_2)(J_1 - J_2) \tag{51}$$

Determining the efficiency of the cycle, as usual, in the form $\eta = \frac{w}{\Delta Q_{12}}$, we obtain for the Carnot cycle

$$\eta = 1 - \frac{\tau_2}{\tau_1} \tag{52}$$

or taking into account (19)

$$\eta = 1 - \frac{\omega_2}{\omega_1} \tag{53}$$

Thus, the efficiency of the Carnot cycle for a Hamiltonian system is determined only by the ratio of the maximum and minimum (over the entire thermodynamic cycle) frequencies of oscillations of the system and does not depend on the type of the Hamiltonian system.

It is easy to show that, for Hamiltonian systems, the efficiency of any other thermodynamic cycle that occurs in the same frequency range $\omega_2 \leq \omega \leq \omega_1$ cannot exceed the efficiency of the Carnot cycle.

## 6 Thermodynamic interaction of Hamiltonian systems

Consider two interacting Hamiltonian systems. We assume that these systems slowly (according to condition (5)) exchange energy, i.e. part of the energy of one system is somehow taken away and completely (without losses) is transferred to another system. According to the definition given above, such an interaction of two Hamiltonian systems is a thermodynamic process. Further, we assume that the external parameters of interacting systems in this process remain unchanged. In this case, according to the above definition, only thermal (in a generalized sense) interaction (heat exchange) takes place between the systems.

We will not consider specific mechanisms of interaction between two Hamiltonian systems, since from the point of view of Hamiltonian thermodynamics, it does not matter; what matters is only the amount of energy transferred from one system to another at each act of interaction, and how often these acts of interaction occur.

We introduce the concept of thermodynamic equilibrium of interacting systems, by which, as usual, we mean a stable state in which the temperatures of the interacting systems are the same. Since the Hamiltonian temperatures for the systems under consideration coincide with the frequencies of their oscillations, then the thermodynamic equilibrium of two Hamiltonian systems means synchronization of their oscillations, at which the oscillation frequencies of the interacting systems coincide.

We can pose the question: under what conditions will two interacting Hamiltonian systems come to a thermodynamic equilibrium state? In other words, under what conditions will synchronization of oscillations of two interacting Hamiltonian systems occur?

Let $E_i$ and $\tau_i$ be the energy and temperature of the $i$th Hamiltonian system; $i = 1,2$; $\Delta Q_{ij}$ is the amount of energy (heat) transferred from system $j$ to system $i$ in one act of interaction. It is considered that $\Delta Q_{ij} > 0$ if system $j$ loses energy (heat), while system $i$ receives it. By virtue of the law of conservation of energy $\Delta Q_{ij} = -\Delta Q_{ji}$.

Considering the thermodynamic interaction of two Hamiltonian systems, we assume for definiteness that $\tau_1 > \tau_2$.

Obviously, in order for these two systems to come to a state of thermodynamic equilibrium, the temperature of system 1 must decrease in the course of interaction, while the temperature of system 2 must increase until they become equal, i.e. until thermodynamic equilibrium $\tau_1 = \tau_2$ (synchronization of oscillations) occurs. In this case, the thermodynamic equilibrium state of the two Hamiltonian systems is stable.

A change in the Hamiltonian temperatures of interacting systems occurs due to a change in their energies.

In this case, one can write

$$dE_1 = -dE_2 = \Delta Q_{12} \qquad (54)$$

As in ordinary thermodynamics, for a Hamiltonian system, one can introduce the heat capacity $C = \frac{\delta Q}{dt}$, which is not a function of state, but depends on the type of process. In particular, for a process that occurs at constant external parameters $A$, the Hamiltonian heat capacity is

$$C_A = \left(\frac{\partial E}{\partial \tau}\right)_A \qquad (55)$$

The Hamiltonian heat capacity $C_A$ is analogous to the heat capacity $C_V$ at constant volume in conventional thermodynamics.

Taking into account the equation of state of the Hamiltonian system (20) and relation (55), we can write equation (54) in the form

$$C_{A1} d\tau_1 = -C_{A2} d\tau_2 = \Delta Q_{12} \qquad (56)$$

If the considered interacting systems tend to the thermodynamic equilibrium state, then $d\tau_1 < 0$ and $d\tau_2 > 0$.

In this case, from equation (56), we obtain the conditions

$$\frac{\Delta Q_{12}}{C_{A1}} < 0, \frac{\Delta Q_{12}}{C_{A2}} < 0 \qquad (57)$$

In contrast to the heat capacity of molecular systems, which is always positive, the Hamiltonian heat capacity (54) can be either positive or negative.

It follows from conditions (57) that interacting systems can reach a thermodynamic equilibrium state only when the heat capacities of both systems have the same sign, i.e. when $C_{A1} C_{A2} > 0$. If the heat capacities of interacting systems have different signs ($C_{A1} C_{A2} < 0$), then the thermodynamic equilibrium of such systems is impossible.

We consider two cases.

(i) $C_{A1} > 0$ and $C_{A2} > 0$. This means that an increase in the energy of each of the systems under consideration leads to an increase in their Hamiltonian temperature and vice versa. In this case, the interacting systems will come to a thermodynamic equilibrium state if

$$\Delta Q_{12} = -\Delta Q_{21} = -\Phi(\tau_1, \tau_2) \qquad (58)$$

where

$$\Phi(\tau_1, \tau_2) = -\Phi(\tau_2, \tau_1); \Phi(\tau_1, \tau_2) > 0 \text{ for } \tau_1 > \tau_2 \text{ and } \Phi(\tau_1, \tau_2) = 0 \text{ for } \tau_1 = \tau_2 \qquad (59)$$

In this case, heat is transferred from a system with a higher Hamiltonian temperature to a system with a lower Hamiltonian temperature, as in ordinary thermodynamics.

Special cases of (58) and (59) are

$$\Phi(\tau_1, \tau_2) = \alpha(\tau_1, \tau_2)(F(\tau_1) - F(\tau_2)) \qquad (60)$$

or even

$$\Phi(\tau_1, \tau_2) = \alpha(\tau_1, \tau_2)(\tau_1 - \tau_2) \qquad (61)$$

where $F(\tau)$ is the monotonically increasing function ($\frac{dF}{d\tau} > 0$); $\alpha(\tau_1, \tau_2) = \alpha(\tau_2, \tau_1) > 0$.

(ii) $C_{A1} < 0$ and $C_{A2} < 0$. This means that an increase in the energy of each of the systems under consideration leads to a decrease in their Hamiltonian temperature and vice versa. In this case, the interacting systems will come to a state of thermodynamic equilibrium if

$$\Delta Q_{12} = -\Delta Q_{21} = \Phi(\tau_1, \tau_2) \qquad (62)$$

where the function $\Phi(\tau_1, \tau_2)$ satisfies conditions (59), and in special cases it can have the form (60) and (61). In this case, in contrast to conventional thermodynamics, heat is transferred from a system with a lower Hamiltonian temperature to a system with a higher Hamiltonian temperature.

Note that the function $\Phi(\tau_1, \tau_2)$ satisfying conditions (59) can be both deterministic and random at each act of interaction of Hamiltonian systems.

If the heat capacity of one of the systems is much higher than the other, for example, $C_{A1} \gg C_{A2}$, then, taking into account (56), we obtain $|d\tau_1| \ll |d\tau_2|$. In particular, at $C_{A1} \to \infty$, the temperature $\tau_1$ will remain constant during the interaction, while the temperature $\tau_2$ will tend to $\tau_1$. In this case, system 1 is a Hamiltonian thermostat, the temperature (frequency of oscillation) of which does not change when interacting with other systems.

Consider two Hamiltonian systems having $E_1 = E_1(J_1, A_1)$ and $E_2 = E_2(J_2, A_2)$, and satisfying the condition $C_{A1} C_{A2} > 0$.

Consider a process in which each of the systems receives energy from the outside in the form of heat (respectively, $\Delta Q_1$ and $\Delta Q_2$), while the external parameters of these systems remain unchanged: $A_1 = const$ and $A_2 = const$ (i.e., no work is done).

We assume that quasi-static interaction (58) or (62) taking into account (59) takes place between the systems, as a result of which they exchange energy in the form of heat $\Delta Q_{12} = -\Delta Q_{21}$ (Fig. 2).

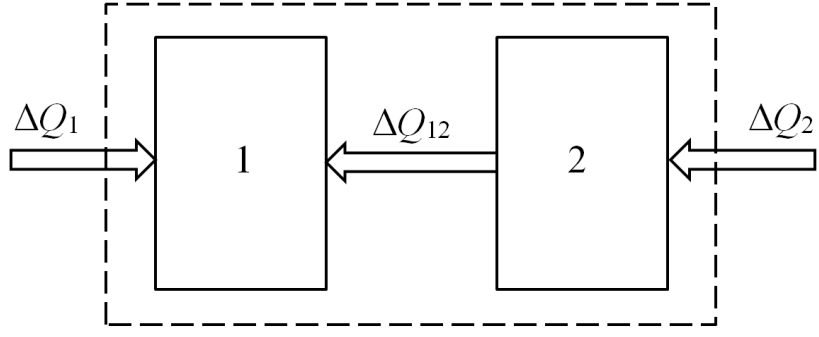

Fig. 2. Interaction of two Hamiltonian systems

For each system under consideration, we can write the second law of thermodynamics (17) which takes the form

$$\Delta Q_{12} + \Delta Q_1 = \tau_1 dJ_1 \qquad (63)$$

$$-\Delta Q_{12} + \Delta Q_2 = \tau_2 dJ_2 \qquad (64)$$

These systems can be considered as subsystems of the combined system (Fig. 2), which has energy $E = E_1 + E_2$, and receives energy from the outside in the form of heat

$$dE = \delta Q = \Delta Q_1 + \Delta Q_2 \qquad (65)$$

Taking into account (63) and (64), one obtains

$$\delta Q = \tau_1 dJ_1 + \tau_2 dJ_2 \qquad (66)$$

If these systems are in thermodynamic equilibrium then

$$\tau_1 = \tau_2 = \tau \qquad (67)$$

In this case, for the combined system as a whole, the second law of thermodynamics (17) holds:

$$\delta Q = \tau dJ \qquad (68)$$

where

$$J = J_1 + J_2 \qquad (69)$$

is the Hamiltonian entropy of the combined system.

Consider a special case

$$\Delta Q_1 = \Delta Q_2 = 0 \qquad (70)$$

when each of the subsystems of the combined system (and, hence, the combined system as a whole) does not receive energy from the outside in the form of heat. In this case, the combined system as a whole is adiabatically isolated, although energy exchange in the form of heat is possible inside it (between its parts).

In this case, from (63) and (64) we obtain $dJ_1 = \Delta Q_{12}/\tau_1$ and $dJ_2 = -\Delta Q_{12}/\tau_2$. Then

$$dJ = \left(\frac{1}{\tau_1} - \frac{1}{\tau_2}\right) \Delta Q_{12} \qquad (71)$$

In this case, depending on the sign of $C_{A1}$ and $C_{A2}$, we obtain:

(i) If $C_{A1} > 0$ and $C_{A2} > 0$, then, taking into account (58) and (59), one obtains

$$dJ \geq 0 \tag{72}$$

where the equal sign takes place only at thermodynamic equilibrium (67) of the considered systems. Thus, in this case, the entropy $J$ of the combined system increases and reaches a maximum at thermodynamic equilibrium. This means that the entropy of the combined system can be considered as the Lyapunov function of this system, such that for an adiabatic system $\frac{dJ}{dt} \geq 0$.

(ii) If $C_{A1} < 0$ and $C_{A2} < 0$, then, taking into account (62) and (59), one obtains

$$dJ \leq 0 \tag{73}$$

where the equal sign takes place only at thermodynamic equilibrium (67) of the considered systems. Thus, in this case, the entropy $J$ of the combined system decreases and reaches a minimum at thermodynamic equilibrium. This means that the entropy of the combined system can be considered as the Lyapunov function of this system, such that for the adiabatic system $\frac{dJ}{dt} \leq 0$.

Equations (72) and (73) represent the second law of thermodynamics for an adiabatically isolated system consisting of two Hamiltonian systems interacting according to laws (58) or (62) taking into account (59).

Inequalities (72) and (73) show that the equilibrium state of such a combined system can be found as a conditional extremum (maximum or minimum) of entropy (69) of the combined system for a given energy $E = E_1(J_1, A_1) + E_2(J_2, A_2)$ of the combined system. This conditional extremum can be easily found by the method of indefinite Lagrange multipliers by introducing the Lagrange function $F(J_1, J_2) = J + \lambda E$, where $\lambda$ is an indefinite Lagrange multiplier. Then the equilibrium state of the system (the conditional extremum of the entropy of the combined system) corresponds to the condition $\frac{\partial F}{\partial J_1} = \frac{\partial F}{\partial J_2} = 0$. Taking into account (14) and (69), we obtain $1 + \lambda \tau_1 = 1 + \lambda \tau_2 = 0$. From this, the condition of thermodynamic equilibrium (67) of the considered Hamiltonian systems follows, where $\tau = -1/\lambda$ is the equilibrium Hamiltonian temperature, which is found from the given energy of the combined system $E_1(\tau, A_1) + E_2(\tau, A_2) = E$, where $E_i(\tau, A_i)$ is the energy equation of state for the $i$th Hamiltonian system; $i = 1,2$.

Now consider a nonadiabatic combined system for which condition (70) is not satisfied. In this case, the energy of the combined system changes according to equation (65). Taking into account (63), (64) and (69), we obtain

$$dJ = \left(\frac{1}{\tau_1} - \frac{1}{\tau_2}\right)\Delta Q_{12} + \frac{\Delta Q_1}{\tau_1} + \frac{\Delta Q_2}{\tau_2} \tag{74}$$

As shown above, for the considered law of interaction of Hamiltonian systems, the term $\left(\frac{1}{\tau_1} - \frac{1}{\tau_2}\right)\Delta Q_{12}$ always has a definite sign, therefore:

(i) If $C_{A1} > 0$ and $C_{A2} > 0$, then, taking into account (58) and (59), one obtains

$$dJ \geq \frac{\Delta Q_1}{\tau_1} + \frac{\Delta Q_2}{\tau_2} \tag{75}$$

where the equal sign takes place only at thermodynamic equilibrium (67) of the considered systems.

If the interacting Hamiltonian systems are in a state close to thermodynamic equilibrium (67), then, taking into account (65), inequality (75) can be written in the form

$$dJ \geq \frac{\delta Q}{\tau} \tag{76}$$

(ii) If $C_{A1} < 0$ and $C_{A2} < 0$, then, taking into account (62) and (59), one obtains

$$dJ \leq \frac{\Delta Q_1}{\tau_1} + \frac{\Delta Q_2}{\tau_2} \tag{77}$$

where the equal sign takes place only at thermodynamic equilibrium (67) of the considered systems.

If the interacting Hamiltonian systems are in a state close to thermodynamic equilibrium (67), then, taking into account (65), inequality (77) can be written in the form

$$dJ \leq \frac{\delta Q}{\tau} \tag{78}$$

Equations (75) and (77) represent the general form of the second law of thermodynamics for a combined system consisting of two Hamiltonian systems interacting according to laws (58) or (62) taking into account (59) and exchanging energy in the form of heat with external systems.

Note that the second law of thermodynamics in the traditional form (76) holds only for the case when $C_{A1} > 0$ and $C_{A2} > 0$.

A special case is when $\Delta Q_2 = -\Delta Q_1$. In this case $dE = \delta Q = 0$, i.e. the energy of the combined system does not change, while the heat is simply "pumped" through the combined system: heat in the amount $\Delta Q_1$ is supplied to system 1 and is immediately taken away from system 2 in the same amount.

In this case, conditions (75) and (77) take, respectively, the form

$$dJ \geq \left(\frac{1}{\tau_1} - \frac{1}{\tau_2}\right)\Delta Q_1 \tag{79}$$

$$dJ \leq \left(\frac{1}{\tau_1} - \frac{1}{\tau_2}\right)\Delta Q_1 \tag{80}$$

In the general case, the sign of $\Delta Q_1$ is in no way related to the sign of $\left(\frac{1}{\tau_1} - \frac{1}{\tau_2}\right)$, therefore nothing definite can be said about the sign of the right-hand side of (79) and (80). However, it is easy to see that in such a process the temperatures of interacting Hamiltonian systems will be changed until equilibrium $dJ_1 = dJ_2 = 0$ occurs. In this case, as follows from (63) and (64), $\Delta Q_1 = \Delta Q_{21} = -\Delta Q_2$, and the heat supplied to system 1 is immediately transferred in full to system 2, from which it is completely taken away. Such an equilibrium is possible only at certain temperatures $\tau_1$ and $\tau_2$, depending on the law $\Phi(\tau_1, \tau_2)$ of thermodynamic interaction of Hamiltonian systems and the energy of the combined system $E$. In thermodynamics, such stationary states of the combined system are called nonequilibrium.

## 7 Concluding remarks

In this paper, we propose a thermodynamic approach to description of slow–fast Hamiltonian systems [5-8].

We have shown that even for a deterministic Hamiltonian system, it is possible to construct thermodynamics similar to ordinary (molecular) thermodynamics. In particular, thermodynamic concepts such as temperature, heat, entropy, etc., can be introduced for a Hamiltonian system. For a deterministic Hamiltonian system, the first and second laws of thermodynamics are obtained in a natural way and are completely analogous to the laws of ordinary (molecular) thermodynamics, but are not related to thermal or other random motion of the system or its parts.

The developed approach allows giving a new thermodynamic formulation of the theory of adiabatic invariants [1-4].

The analysis performed allows a deeper understanding of the foundations of conventional (molecular) thermodynamics and statistical mechanics.

In particular, we can conclude, if the thermodynamic temperature (i.e., the random force acting to the system) tends to zero, then the system under consideration will turn from thermodynamic into a deterministic Hamiltonian system, which can be described with Hamiltonian thermodynamics considered in this paper. In other words, with the cessation of random motion, thermodynamics does not "disappear", but passes into Hamiltonian thermodynamics. Moreover, as shown in this paper, it is possible to realize the thermodynamic processes in a generalized sense (as an energy exchange between fast and slow degrees of freedom) even at zero thermodynamic temperature. Hence it follows that the "complete" thermodynamics of the system should include both conventional (molecular, stochastic) thermodynamics and Hamiltonian thermodynamics, which is currently not taken into account. This conclusion seems to be important, especially when applied to quantum systems, taking into account the rapidly growing interest in quantum thermodynamics [9].

Note that the developed theory is applicable not only to mechanical systems, but also to any Hamiltonian systems, regardless of their nature.

The thermodynamics of more complex near-Hamiltonian systems as well as the general thermodynamic theory of dynamical systems will be considered in the next papers of this series.

**Acknowledgments**

This work was done on the theme of the State Task No. AAAA-A20-120011690135-5. Funding was provided in part by the Tomsk State University competitiveness improvement program.

**Conflict of Interest:** The author declares that he has no conflict of interest.